\documentstyle[12pt]{article}

\newcommand{\beq}{\begin{equation}}
\newcommand{\eeq}{\end{equation}}
\newcommand{\beqa}{\begin{eqnarray}}
\newcommand{\eeqa}{\end{eqnarray}}

\begin{document}

\begin{flushright}
CERN-TH/97-134\\
June 1997\\
hep-ph/9706457
\end{flushright}

\vspace*{1cm}
\begin{center}
{\Large\bf Renormalon Phenomenology: Questions and Directions}
\footnote{Talk presented at the Fifth International Workshop on 
`Deep Inelastic Scattering and QCD' (DIS'97), Chicago, IL, April 1997.}
\end{center}

\vspace{0.7cm}
\begin{center}
{\sc M. Beneke}\\[0.5cm]
{\sl Theory Division, CERN, CH-1211 Geneva 23, Switzerland}\\[0.3cm]
\end{center}

\vspace*{1cm}
\begin{abstract} 
\noindent A qualitative 
(and selective) discussion of current activities and 
problems in the field is given. 
\end{abstract}
\vspace*{1cm}

\noindent The basic idea of `renormalon phenomenology' is simply this: to 
parametrise, or at least unravel, power-like infrared (IR) sensitive 
contributions to hard processes. Just as perturbative IR 
logarithms in QCD lead to the introduction of non-perturbative parameters, 
such as parton densities in DIS processes, so do power-like 
dependences on an IR factorisation scale indicate power-suppressed 
non-perturbative contributions. In DIS they are known as higher-twist 
corrections. Now one aims at a more general understanding of power 
corrections, including truly Minkowskian processes. The fact that this 
is connected with renormalons, that is, large-order behaviour in 
perturbation theory, can be considered as an accident of history. 
[The connection between large orders and small momentum 
is by itself quite interesting and has led to a 
better understanding of the systematics of exact multi-loop results. 
This, however, is not the subject of the present talk.]
In a wider sense the problem is the generalisation of perturbative 
factorisation beyond leading power accuracy. 

DIS structure functions provide a useful example to start with. The 
twist expansion of the longitudinal structure function can be written 
as
\begin{eqnarray}
\label{fl}
F_L(x,Q)/(2 x) 
&=& \sum_i\int\limits_x^1\frac{d\xi}{\xi}\,C_i^{[2]}(x,\xi,Q,\mu)
\,f_i(\xi,\mu)
\nonumber\\
&&\,+\frac{1}{Q^2}\sum_j\int \!d\xi_1 d\xi_2\,C_j^{[4]}(x,\xi_1,\xi_2,Q,\mu)
\,T_j(\xi_1,\xi_2,\mu) + \ldots.
\end{eqnarray}
It is well-known that due to logarithmic operator mixing, the factorisation 
scale dependence in the leading twist term cancels only over different 
contributions (quarks and gluons) in the sum over $i$. It is less known 
that the separation of twist-2 and twist-4 is also not unique. It 
would be obvious, if the factorisation in transverse momentum were 
introduced explicitly, in which case powers of $\mu^2/Q^2$ would arise. 
In dimensional regularisation, the ambiguity in separating twists appears 
as renormalons: the series expansion of the coefficient function $C^{[2]}_i$ 
diverges in large orders in $\alpha_s$ and can not be unambiguously 
defined. This IR renormalon divergence comes from small momenta in the 
loops. The ambiguity is compensated by corresponding ultraviolet 
contributions to the matrix elements $T_j$ of twist-4 (non-local) 
operators. This point is crucial: although IR renormalons are IR compared 
to the scale $Q$, they correspond to ultraviolet effects viewed from the 
scale $\Lambda$, the scale of QCD. As a consequence, we can learn very 
little on the specifics of non-perturbative effects. What one does learn 
is the scaling of power corrections with the scale $Q$, just from the 
consistency requirement that a physical quantity must be unambiguous.

Calculations rely on approximations and to date these correspond to an 
analysis of IR sensitive regions in one-loop virtual corrections or 
one gluon emission. The set of bubble diagrams is one way to trace these 
regions through the large-order behaviour of these diagrams \cite{Z}. 
The same set of graphs can 
also be evaluated through a dispersion relation for the running 
coupling \cite{BB1,DMW}. The IR contributions can then be found as 
non-analytic terms at small values of the dispersion variable. 
For sufficiently inclusive observables the 
dispersion variable can be identified with a gluon mass \cite{BBZ}. For other 
interesting quantities like event shape variables and fragmentation 
functions calculations with a gluons mass correspond to an alternative 
scheme of IR regularisation, not related to renormalons.

The past two years have seen many phenomenological applications of 
renormalons. Usually they involve the type of calculations just described, 
together with additional assumptions that can be judged only by their 
empirical success. The second line of interest pursues `operator 
interpretations' of renormalons in Minkowski space, in analogy to 
higher-twist operators in DIS. If successful, one could then dispense 
of particular sets of diagrams and approach an understanding of power 
corrections comparable to applications of operator product expansions 
in Minkowski space. The following gives a telegram overview of some of the 
problems that have been addressed recently.\\ 

{\em The Drell-Yan process.} This process has been the first process 
without operator product expansion, where power corrections have been 
analysed with the help of renormalons. Since collinear factorisation 
can be extended to $1/Q^2$ corrections \cite{QIU91}, the main question 
is whether soft gluons could invalidate this result and introduce 
$1/Q$ corrections. [$Q$ is the mass of the Drell-Yan pair.] The Drell-Yan 
process seems to be well-suited to address this question, as soft 
gluon radiation has been extensively studied and the resummation of 
corresponding large logarithms is well understood.

The first investigations \cite{DY1} of renormalons in Drell-Yan production 
accordingly started 
from the soft gluon resummation formula and reported the presence of $1/Q$ 
corrections. It was then realized \cite{BBDY} that the approximations 
legitimate for a systematic resummation of logarithms lose $1/Q$ power 
corrections and that $1/Q$ corrections cancel in the full result, when 
these approximations are abandoned. While the leading logarithms originate 
from the region $k_\perp\ll k_0\ll Q$, the region 
$k_\perp\sim k_0$, that is, large angle gluon radiation, is equally 
important for power corrections. 

The result of \cite{BBDY} is based on the analysis of one-loop diagrams. 
Although the absence of $1/Q$ corrections to 
all orders may be plausible, this has been shown so far only in 
an abelian theory \cite{ASZ}, where it follows from the fact that 
one gluon emission is already the only building block for multiple soft 
gluon emission amplitudes. For QCD, a proof is still missing. Since the 
non-abelian vertices enter only at two loops, it might be useful  
to extend the analysis of IR sensitive regions to two-loop diagrams. 
This would also provide a check on possible interpretations of power 
corrections to Drell-Yan production in terms of some operators. If 
$1/Q$ corrections are indeed absent, as we believe,  the same twist-4 
multi-parton correlations that enter DIS would seem the best bet for these 
operators at order $1/Q^2$ \cite{QIU91}.\\

{\em Hadronic event shape variables}\cite{E1,E2,E3}. These are the 
simplest observables for which $1/Q$ corrections have been found. They 
come only from the soft region and it is natural to associate them 
with hadronization corrections. In addition, experimental information exists 
for various center-of-mass energies $Q$ in $e^+ e^-$ collisions and a
$1/Q$ term fits the difference between data and NLO perturbation very well 
\cite{delphi}. 

One can take a step further towards the absolute magnitude of power 
corrections by assuming that $1/Q$ corrections are universal \cite{E2}, 
i.e. that a single non-perturbative number controls $1/Q$ corrections 
to all event shape variables. To be precise, one assumes that for 
any (averaged) event shape $S$ (such as the average $1-T$ etc.), the 
$1/Q$ power correction is given by
\begin{equation}
\label{s} 
S_{|1/Q} = K_S\cdot\frac{\langle \mu_{had}\rangle}{Q}
\end{equation}
with a unique parameter $\langle \mu_{had}\rangle$ and a calculable 
coefficient $K_S$ that depends on $S$. Note that if one 
identifies $K_S$ with the residue of the IR renormalon pole that 
leads to the $1/Q$ ambiguity, $K_S$ can not be calculated, because 
all higher order diagrams contribute to the residue. However, the 
universality assumption implies that they contribute equally to different 
event shapes so that the ratio $K_{S_1}/K_{S_2}$ can be found by a 
one-loop calculation up to corrections of order $\alpha_s(Q)$. 
Since one event shape measurement is required to fix 
$\langle \mu_{had}\rangle$, knowledge of the ratio is sufficient to 
predict $1/Q$ corrections to other event shapes.

The universality assumption seems to work well phenomenologically, 
to an accuracy of about 20\%. The calculation has now been done 
also for event shape variables in DIS \cite{DWDIS} and a good 
fit to the data 
is obtained with the same value of 
$\langle \mu_{had}\rangle$ as in $e^+ e^-$ collisions \cite{EDIS}. 
This could not have been expected theoretically.

It is fair to say that even for event shape variables in 
$e^+ e^-$ collisions alone a good theoretical 
argument in favour of universality still remains to be found. 
Diagrammatically it is evident that higher order contributions do 
not contribute equally to the renormalon residue \cite{E3}, mainly 
because event shapes resolve large angle soft gluon emission at the 
level of $1/Q$ power corrections even in the two-jet limit \cite{BBDY}. 
As a consequence every event shape corresponds to a different weight on 
the distribution of soft gluons \cite{sterman} (see also \cite{BBM}), 
which can not be described by a single number $\langle \mu_{had}\rangle$. 
Another problem is the `uniqueness problem': even if universality held 
(or held approximately), we do not know how to calculate ratios 
$K_{S_1}/K_{S_2}$ unambiguously. As mentioned above, calculations based 
on bubble diagrams or on a finite gluon mass as IR regulator lead to 
different results \cite{BBDY,E3,BBM}. Since there is no obvious reason to 
prefer one or the other, this difference must be considered as an 
uncertainty in $K_{S_1}/K_{S_2}$. For the longitudinal cross section, 
this difference is small \cite{BBM}, about 20\%. Given these 
reservations, the fact that universality appears to work 
approximately is by itself quite interesting and remains to be 
understood.\\

{\em $x$-dependence of twist-4 corrections in DIS} \cite{DMW,x}. Since 
the moments of DIS structure functions have an operator product 
expansion, one may not expect renormalons to give any further insight on 
their power behaviour.  
However, if one assumes that the $x$-dependence of the twist-4 correction 
follows the $x$-dependence of the corresponding IR sensitive contribution 
in perturbation theory, the `unknowns' at twist-4 are reduced from 
functions to numbers. This is a very strong assumption and one may 
consider the result as a model for twist-4 multi-parton correlation 
functions. For example, eq.~(\ref{fl}) reduces to   
\begin{equation}
\label{fl2}
F_L(x,Q)/(2 x) 
= \sum_i\int\limits_x^1\frac{d\xi}{\xi}\,f_i(x/\xi,\mu)\,
\left[C_i^{[2]}(\xi,Q,\mu)+A_i(\xi)\frac{\Lambda^2}{Q^2}\right]
+ \ldots
\end{equation}
with calculable functions $A_i(\xi)$. Comparison with twist-4 corrections 
extracted from DIS data shows surprising agreement of the model with the 
$x$-dependence of the data. 

 Theoretically it seems rather obscure that the model should work so well 
and again it is the data itself that teaches us an interesting lesson. Two 
suggestions have been put forward: the first \cite{DMW} is based on the 
idea of universality, which postulates that all non-perturbative effects 
are generated through integrals over an IR finite coupling, at least to 
first approximation. This postulate 
itself appears hard to digest; the second \cite{BBM} is an 
a posteriori explanation based on the correspondence of IR renormalons 
and cut-off dependence of higher-twist operators: the above model could be 
justified, if the matrix elements of twist-four operators were 
dominated by their cut-off piece rather than by `genuine' non-perturbative 
effects. Again, there is no dynamical understanding why this 
should be so, unless we consider both $A_i(\xi)$ and the data on 
twist-4 corrections as effective parametrisation of higher order 
perturbative corrections beyond NLO. Another possibility is that 
one is mainly seeing generic $x$-dependences that follow from kinematics 
such as $A_i(x)/C_i^{[2]}(x)\sim 1/(1-x)$ as $x\to 1$. 

 Whatever the agreement with the data, the model by construction 
gives no insight into 
hadron structure. In other words, in terms of moments
\begin{equation}
\label{moment}
\frac{M_n^{twist-4}}{M_n^{twist-2}}|_{\mbox{hadron 1}} - 
\frac{M_n^{twist-4}}{M_n^{twist-2}}|_{\mbox{hadron 2}} \equiv 0.
\end{equation}
The model can only work for those aspects of twist-4 corrections which 
are not hadron-specific. Let me also note that in present applications, 
the added twist-4 correction does not have the correct scale-dependence. 
This constitutes an additional uncertainty, although the problem is 
not fundamental and could be solved by higher-loop calculations of 
$A^i(\xi)$.\\

{\em Fragmentation} \cite{BBM,DW}. The same model has been suggested 
for fragmentation processes in $e^+ e^-$ annihilation, although there 
is as yet no data to compare with. The uniqueness problem mentioned 
for event shape variables affects the calculation of the functions 
$A^i(x)$ in fragmentation even more severely. On the other hand 
fragmentation processes are particularly interesting, because although 
the leading-twist formalism is analogous to DIS, there is no operator 
product expansion to provide information beyond leading power. It has 
been found that for non-zero value of the scaling variable $x$, the 
leading power corrections are of order $1/Q^2$, in agreement with 
the collinear expansion of fragmentation functions in \cite{BBOPE}. 
But this statement ceases 
to be true for moments in $x$, which include the soft region at 
small $x$. The power expansion of fragmentation 
processes has strong soft gluon singularities at small $x$, which are 
non-integrable. After integration over $x$, including the small-$x$ 
region, every term $(\Lambda^2/(Q^2 x^2))^n$ in the expansion at 
non-zero $x$ can contribute to the leading power correction to the 
moments. The leading power correction depends on the order of the 
moment, a situation that is excluded for moments of DIS structure 
functions. In particular, the total longitudinal cross section receives 
a $1/Q$ power corrections similar to other event shapes. Only for 
high enough moments is the soft region suppressed and power corrections 
scale universally as $1/Q^2$ due to collinear regions. As a consequence, 
a light-cone expansion for moments of fragmentation expansions does 
not exist. Note that in DIS and fragmentation, the twist expansion 
also breaks down as $x\to 1$. However, the singularities as $x\to 1$ 
are integrable and do not alter the power behaviour of the power 
expansion of moments. 

It is known that at small $x$ multiple gluons emission and coherence 
effects lead to a suppression of soft hadron production, which formally 
follows from resummation of logarithms in $x$ in perturbation theory. 
One may wonder how multiple soft gluon emission would affect the 
$x$-behaviour of power corrections in the small-$x$ region, which has 
been crucial in understanding the emergence of a $1/Q$ correction to 
the longitudinal cross section. It may well be that the two problems 
are in fact disconnected because they refer to different momentum regions 
in Feynman integrals. On the other hand, the question of how to reconcile 
the two different sets of higher order diagrams that correspond to 
multiple gluon emission and to renormalons has not yet been addressed.\\

{\em Soft gluon cancellations beyond logarithms.} A more general 
understanding of under which circumstances $1/Q$ corrections exist 
would be highly desirable. From what we have learnt, such corrections, 
if present, would arise only from soft, but not from collinear 
partons. The approach to the problem taken in \cite{AZ} makes essential 
use of the KLN and Low theorems, but still has to be completed 
for the non-abelian theory. As 
shown in \cite{AZ} the KLN theorem alone already guarantees the 
absence of $1/Q$ corrections to a process inclusive over degenerate 
final and initial states, basically because the KLN transition 
amplitudes have no $1/k^0$ contributions (where $k^0$ is the emitted 
gluon's energy). As a consequence, the amplitude squared integrated 
over phase space is proportional to $dk^0 k^0$ for small $k^0$ which 
implies at most $1/Q^2$ corrections. The main obstacle in this 
approach appears 
to be the generalisation of the Low theorem required to dispense of the 
sum over degenerate initial states implied by the KLN theorem. 

Another possibility would be to investigate 
in what situations the Ward identities that guarantee the cancellation 
of soft gluon divergences in the factorisation proofs generalise to 
the cancellation of $1/Q$ corrections as well.\\ 

More theoretical work is needed particularly 
in this direction. Meanwhile, the 
phenomenology of renormalons is encouraging, but some magic seems to 
be at work: things work that needn't have to.

\end{document}